\documentstyle[12pt]{article}

\def\a{\alpha}

\def\d{\delta}

\def\h{\eta}

\def\k{\kappa}                  

\def\m{\mu}

\def\p{\pi}                     
\def\t{\tau}

\def\F{\Phi}

\def\J{\Psi}

\def\P{\Pi}

\def\ca{{\cal A}}

\def\cs{{\cal S}}

\newcommand{\extraspace}{\addtolength{\abovedisplayskip}{2mm}
                        \addtolength{\belowdisplayskip}{2mm}
                        \addtolength{\abovedisplayshortskip}{2mm}
                        \addtolength{\belowdisplayshortskip}{2mm}}
\newcommand{\be}{\begin{equation}\extraspace}

\newcommand{\ee}{\end{equation}}

\newcommand{\bea}{\begin{eqnarray}\extraspace}
\newcommand{\beastar}{\begin{eqnarray*}\extraspace}
\newcommand{\eea}{\end{eqnarray}}
\newcommand{\eeastar}{\end{eqnarray*}}

\newcommand{\nonu}{\nonumber \\[2mm]}

\newcommand{\strutje}{\rule[-1.5mm]{0mm}{5mm}}

\newcommand{\half}{\frac{1}{2}}

\newcommand{\wha}{\widehat{\cal A}}

\newcommand{\del}{\partial}

\newcommand{\bdel}{\bar{\partial}}

\newcommand{\bz}{\bar{z}}
\newcommand{\slt}{sl(2)}

\newcommand{\np}{Nucl.\ Phys.\ }

\newcommand{\cmp}{Comm.\ Math.\ Phys.\ }
\newcommand{\pl}{Phys.\ Lett.\ }

\catcode`\@=11
\def\@afterindentfalse{\let\if@afterindent=\iftrue}
\catcode`\@=12

\begin{document}
\font  \biggbold=cmbx10 scaled\magstep2
\font  \bigbold=cmbx10 at 12.5pt
\font  \bigreg=cmr10 at 12pt

\bigreg

\baselineskip = 15pt

\noindent June 1993 \hfill LBL-34125, UCB-PTH-93/19\\
$\strutje$ \hfill  KUL-TF-93/21\\
$\strutje$ \hfill  hep-th/yymmmdd\\

\vspace{4mm}

\begin{center}
{\large EXTENSIONS OF THE VIRASORO ALGEBRA\\ AND GAUGED WZW
MODELS\footnote{This work was
supported in part by the Director,
Office of Energy Research, Office of High Energy and Nuclear Physics,
Division of High Energy Physics of the U.S. Department of Energy
under Contract DE-AC03-76SF00098 and in part by the National Science
Foundation under grant PHY90-21139.} }\\

\vspace{1cm}

{\bf \centerline{Alexander Sevrin\footnote{Address after October 1st, 1993:
CERN, TH Division, CH-1211 Gen\`eve, Switzerland}${}^1$ and
Walter Troost\footnote{Bevoegdverklaard Navorser NFWO,Belgium}${}^{2}$}}
\vskip .3cm
{\baselineskip = 12pt
\centerline{\sl{1. Department of Physics}}
\centerline{\sl{University of California at Berkeley}}
\centerline{\sl{and}}
\centerline{\sl{Theoretical Physics Group}}
\centerline{\sl{Lawrence Berkeley Laboratory}}
\centerline{\sl{Berkeley, CA 94720, U.S.A.}}}
\vskip .2cm
{\baselineskip = 12pt
\centerline{\sl{2. Instituut voor Theoretische Fysica}}
\centerline{\sl{Universiteit Leuven}}
\centerline{\sl{Celestijnenlaan 200D, B-3001 Leuven, Belgium}}}
\end{center}

\vskip 1.cm
\centerline{\bf Abstract}
\vskip 0.5cm
{\baselineskip=14pt \small
To any non-trivial embedding of $\slt$ in a (super) Lie algebra, one
can associate an extension of the Virasoro algebra. We realize the extended
Virasoro algebra in terms of a WZW model in which a chiral, solvable group is
gauged, the gauge group being determined by the $\slt$ embedding. The resulting
BRST cohomology is computed and the field content of the extended Virasoro
algebra is determined. The closure of the extended Virasoro algebra is shown.
Applications such as the quantum Miura transformation and the effective action
of the associated extended gravity theory are discussed.}

\newpage

\renewcommand{\thepage}{\roman{page}}
\setcounter{page}{2}
\mbox{ }

\vskip 1in

\begin{center}
{\bf Disclaimer}
\end{center}

\vskip .2in

\begin{scriptsize}
\begin{quotation}
This document was prepared as an account of work sponsored by the United
States Government.  Neither the United States Government nor any agency
thereof, nor The Regents of the University of California, nor any of their
employees, makes any warranty, express or implied, or assumes any legal
liability or responsibility for the accuracy, completeness, or usefulness
of any information, apparatus, product, or process disclosed, or represents
that its use would not infringe privately owned rights.  Reference herein
to any specific commercial products process, or service by its trade name,
trademark, manufacturer, or otherwise, does not necessarily constitute or
imply its endorsement, recommendation, or favoring by the United States
Government or any agency thereof, or The Regents of the University of
California.  The views and opinions of authors expressed herein do not
necessarily state or reflect those of the United States Government or any
agency thereof of The Regents of the University of California and shall
not be used for advertising or product endorsement purposes.
\end{quotation}
\end{scriptsize}

\vskip 2in

\begin{center}
\begin{small}
{\it Lawrence Berkeley Laboratory is an equal opportunity employer.}
\end{small}
\end{center}

\newpage
\renewcommand{\thepage}{\arabic{page}}
\setcounter{page}{1}

\baselineskip=17pt

\setcounter{equation}{0}
\setcounter{footnote}{0}

Extensions of the Virasoro algebra, for a review see \cite{pbks}, such as the
$W_n$ algebras, the supersymmetric Virasoro algebra, etc., play a crucial role
in the study of conformal field theories, $2D$ gravity and integrable systems.
There is a close relation between extended Virasoro algebras and embeddings of
$\slt$ in a
(super) Lie algebra $\bar{g}$. In \cite{bais}, it was shown that given an
affine Lie algebra $\hat{g}$ and an embedding of $\slt$ in $\bar{g}$, one
recovers the Ward identities of an extended Virasoro algebra from the Ward
identities of a WZW model by constraining the affine currents as $J_z\equiv
\frac \k 2 e_-+T$ where $e_\pm$ and $e_0$ are the $\slt$ generators and
$T\in\ker \mbox{ad} e_+$.
Almost \cite{kausch} all known extended Virasoro algebras can be obtained
in this way. Some algebras which do not fall into this class can be obtained
from an algebra which falls in this class by orbifolding. Another complication
which arises is that for certain extended
Virasoro algebras, currents of dimension $1/2$ appear. Dimension $1/2$ currents
cannot be obtained by the method outlined above. However because of the fact
that dimension $1/2$ currents can always be decoupled from the extended
Virasoro algebra, \cite{goddard}, see also \cite{kris1}, these cases are
covered as well. Many aspects of these reductions were studied in
\cite{dublin,dublin2}. Until recently, most considerations were classical. In
\cite{frenkel,tjin2} important steps towards the understanding of the quantum
theory were made.

Parallel to this issue was the development of induced and effective gravity
theories (in the light-cone gauge) associated to extended Virasoro algebras
\cite{aleks} - \cite{kris2}.

In this letter we complete the program started in \cite{kris2} which relates
both issues discussed above. For an arbitrary embedding of $\slt$ in a Lie
algebra $\bar{g}$, we realize the corresponding extended Virasoro algebra in
terms of a WZW model  for which a chiral, solvable group has been gauged, the
gauge group being determined by the $\slt$ embedding. Gauge invariance requires
in certain cases the introduction of extra free fields. The generators of the
extended Virasoro algebra are obtained by solving the BRST cohomology. Due to
the presence of the extra, free fields, the BRST charge cannot be decomposed
into two mutually anticommuting, nilpotent charges. So, the techniques of
\cite{frenkel,tjin2} cannot be applied
directly. However, because of the existence of a
filtration of the BRST complex, spectral sequence techniques can still be used
to solve the cohomology. The quantum Miura transformation immediately follows
from this. Finally, choosing a different gauge yields
effortlessly the effective
action in the light-cone gauge of the corresponding extended gravity theory.

Given a (super) affine Lie algebra $\hat{g}$ of level $\k$, we call the finite
dimensional  subalgebra $\bar{g}$. Consider a nontrivial embedding of $\slt$ in
a (super) Lie algebra $\bar{g}$. A thorough study of $\slt$ embeddings can be
found in \cite{dynkin}. The adjoint
representation of $\bar{g}$ branches into irreducible representations of
$\slt$. For a given embedding, we denote the generators of $\bar{g}$ by
$t_{(jm,\a_j)}$ where
$j\in \half{\bf N}$ labels the irreducible representation
of $\slt$, $m$ runs from $-j$ to $j$ and $\a_j$ counts the multiplicity of the
irreducible representation $j$ in the branching. The $\slt$ generators $e_\pm$
and $e_0$ are denoted by $e_\pm\equiv t_{(1\pm 1,0)}/\sqrt{2}$ and $e_0\equiv
t_{(10,0)}$. The $\slt$ algebra is given by $[e_0,e_\pm ]=\pm 2 e_\pm$ and
$[e_+,e_-]=e_0$. The action of the $\slt$ algebra on the other generators is
given by
\bea
[e_0,t_{(jm,\a_j)}]&=&2m\, t_{(jm,\a_j)}\nonu
[e_\pm, t_{(jm,\a_j)}]&=&(-)^{j+m-\frac 1 2 \pm\frac 1 2}\sqrt{(j\mp m)(j\pm
m+1)} t_{(jm\pm 1,\a_j)}
\eea
Throughout the paper we will use projection operators $\P$, {\it e.g.}
$\P_+\bar{g}=\{ t_{(jm,\a_j)}|m > 0 ;\forall j,\a_j\}$,
$\P_{\geq m}\bar{g}=\{ t_{(jn,\a_j)}|n \geq m ;\forall j,\a_j\}$,
$\P_m\bar{g}=\{ t_{(jm,\a_j)}|\forall j,\a_j\}$.
Our conventions are as in \cite{kris2}\footnote{Except: in \cite{kris2},
auxiliary fields $\t$, $r$ and $\bar{r}$ were
introduced. The relation between the auxiliary fields here and those in
\cite{kris2} is $[\t_{\mbox{here}} , e_-]=(\t+r+\bar{r})_{\mbox{there}}$.}.

The affine Lie algebra $\hat{g}$ is realized by a WZW model with action $\k
S^-[g]$. The action $\cs_1$
\be
{\cal S}_1=\k S^-[g]+ \frac{1}{\p x} \int str\, A_{\bz}\left( J_z
-\frac \k 2 e_- -\frac \k 2 [\t,e_-]\right)
+ \frac{\k}{4\p x}\int str [\t,e_-]\bdel\t,
\label{actiongen}
\ee
with the affine currents $J_z=\frac \k 2 \del g g^{-1}$, the gauge fields
$A_{\bz}\in\P_{+}\bar{g}$ and the ``auxiliary''
fields
$\t\in\P_{+1/2}\bar{g}$,
is invariant under
\be
g\rightarrow h g\qquad\quad
A_{\bz}\rightarrow\bdel h h^{-1}+h A_{\bz} h^{-1}\qquad\quad
\t\rightarrow \t + \P_{+\frac 1 2}\h,
\ee
where $h=\exp \h$, $\h\in\P_+\bar{g}$. Note that as bosonic irreducible
representations of half integer spin always occur in pairs, the introduction of
auxiliary fields can be avoided in the purely bosonic case \cite{dublin,alex}
by further restricting the gauge group. In the supersymmetric case however,
this is not true anymore. The simplest example, the standard $N=1$
Neveu-Schwarz algebra, which is based on the embedding of $\slt$ in $osp(2|1)$,
already requires the introduction of one extra free fermion. For uniformity and
simplicity, we always introduce these extra fields whenever representations of
half-integer spin occur.

The gauge fields $A_{\bz}$ are Lagrange multipliers which impose the constraint
$\P_-J_z=\frac\k 2 e_-+\frac \k 2 [\t,e_-]$. Call the constrained current
$J^c_z$. Performing the gauge transformation which brings $J_z^c$ in the form
$T+\frac \k 2 e_-$ where $T\in\ker ad e_+$, we get, by construction, the fields
$T$ which are gauge invariant modulo the constraints, {\it i.e.} modulo the
equations of motion of the gauge fields $A_{\bz}$. They are of the form
$T\propto \P_{\ker\, ad\, e_+}J_z+\cdots$

We couple these currents to sources and modify the action to
\be
{\cal S}_2={\cal S}_1+\frac{1}{4\p x y} \int str \m T,\label{invact}
\ee
with the sources $\m\in \ker ad e_-$.
We will show that $T$ forms an extension of the Virasoro algebra.

As the polynomials are only gauge invariant modulo terms proportional to
the equations of motion of the gauge fields, we cancel the resulting
non-invariance terms in $\d \cs_2$ by modifying the
transformation rules for the gauge fields suitably.
These modifications are proportional to the $\m$-fields and do not depend on
the gauge fields themselves. Because the gauge fields occur linearly in the
action, the action $\cs_2$ is now gauge invariant.
In order to gauge fix the system we use the Batalin-Vilkovisky formalism,
\cite{bv,proeyen}. We introduce ghostfields $c\in\P_+\bar{g}$ and anti-fields
$J^*_z\in\bar{g}$, $A^*_{\bz}\in\P_-\bar{g}$, $\t^*\in\P_{-1/2}\bar{g}$ and
$c^*\in\P_{-}\bar{g}$. The solution to the BV master equation is given by:
\bea
\cs_{\rm BV}&=&\cs_2-\frac{1}{2\p x}\int str c^*cc + \frac{1}{2\p x}\int str
J^*_z\left(\frac \k 2 \del c + [c,J_z] \right)
+ \frac{1}{2\p x}\int str \t^*c\nonu
&&+\frac{1}{2\p x}\int str A^*_{\bz}\left(\bdel c + [c,A_{\bz}]
+\mbox{$\m$-dependent terms}\right).\label{bvbv}
\eea
The $\m$-dependent terms proportional to $A^*_{\bz}$ absorb all
complications arising from the non-invariance of $T$. Determining them requires
an explicit knowledge of the extra terms which were added to the transformation
rule of $A_{\bz}$. We will not need this here.

We now perform a canonical transformation which changes $A^*_{\bz}$ into a
field,
the antighost $b\in\P_-\bar{g}$, and $A_{\bz}$ into an antifield $b^*$. This
amounts to the gauge choice  $A_{\bz}=0$. The gauge-fixed action reads:
\be
{\cal S}_{\rm gf}=\k S^-[g]
+ \frac{\k}{4\p x}\int str [\t,e_-]\bdel\t
+\frac{1}{2\p x}\int str\, b\bdel c+\frac{1}{4\p x y} \int str \m \hat{T},
\label{actiongen2}
\ee
and the BRST charge is:
\be
Q=\frac{1}{4\p i x}\oint str\left\{ c \left( J_z -\frac \k 2
e_--\frac \k 2[\t,e_-]+ \frac 1 2 J_z^{\rm gh}\right) \right\},
\ee
where $J_z^{\rm gh}=\frac 1 2 \{b,c\}$.
It is nilpotent. The total current $\hat{J}_z=J_z+J_z^{\rm gh}$ satisfies the
same operator algebra as $J_z$, except for the central extension. If we write
$\hat{J}_z=t^A\hat{J}_{z A}$, $\P_-\hat{J}_z=t^a\hat{J}_{z a}$ and $({\bf
1}-\P_-)\hat{J}_z=t^{\bar{a}}\hat{J}_{z \bar{a}}$, we find
\bea
\hat{J}_{z a}(x) \hat{J}_{z b}(y)&=&(x-y)^{-1} f_{ab}{}^c\hat{J}_{z c}(y) \nonu
\hat{J}_{z a}(x) \hat{J}_{z \bar{b}}(y)&=&\left(-\frac \k 2 g_{a\bar{b}}+
(-)^{(c)} f_{ca}{}^d f_{d\bar{b}}{}^c\right) (x-y)^{-2}+(x-y)^{-1}
f_{a\bar{b}}{}^C\hat{J}_{z C}(y) \nonu
\hat{J}_{z \bar{a}}(x) \hat{J}_{z \bar{b}}(y)&=&\left(-\frac \k 2
g_{\bar{a}\bar{b}}+(-)^{(c)} f_{c\bar{a}}{}^df_{d\bar{b}}{}^c\right)
(x-y)^{-2}+(x-y)^{-1} f_{\bar{a}\bar{b}}{}^{\bar{c}} \hat{J}_{z \bar{c}}(y).
\eea
The only unknown in the action is the current $\hat{T}$. This again reflects
the fact that we did not specify the explicit form of the $\m$ dependent terms
in eq. (\ref{bvbv}). For $\m=0$ the action is BRST invariant. In order to
guarantee BRST invariance
for $\m\neq 0$, the currents $\hat{T}$ themselves have to be BRST invariant.
This determines them up to BRST exact pieces.

We now study the BRST cohomology in detail. Our methods are inspired by
\cite{frenkel,tjin1,tjin2,bott}. However the analysis in
\cite{frenkel,tjin1,tjin2} is based on the presence of a double complex. As we
will see, the $\t$ fields obstruct the existence of a double complex.
Nevertheless, spectral sequence techniques \cite{bott} are still applicable.

Consider the algebra $\ca$ generated by the basic fields
$\{b,\hat{J}_z,\t,c\}$, which consists of all regularized
products\footnote{We use the standard point-splitting regularization: $(AB)(z)=
\frac{1}{2\p i}\oint dz'(z'-z)^{-1}A(z')B(z)$.} of the basic fields and their
derivatives modulo the usual relations \cite{bbss,spindel}
between different orderings, derivatives, etc.
To every field $\F$, we assign a double grading $[\F]=(k,l)$,
$k,\,l\in\frac 1 2 {\bf Z}$, with $k+l\in {\bf Z}$ the ghostnumber:
$[J_z]=(m,-m)$ for $J_z\in\P_m\bar{g}$, $m\in\frac 1 2 {\bf Z}$, $[b]=(-m,m-1)$
for $b\in\P_{-m}\bar{g}$, $m>0$, $[c]=(m,-m+1)$ for $c\in\P_m\bar{g}$, $m>0$
and $[\t]=(0,0)$. Through this $\ca$ acquires a double grading:
\be
\ca=\bigoplus_{\stackrel{\scriptstyle m,n\in\frac 1 2{\bf Z}}{m+n\in{\bf
Z}}}\ca_{(m,n)}.
\ee
The operator product expansions (OPE) are compatible with the grading.

The idea of spectral sequences is now to compute the cohomology in steps.
One starts by working to 'leading order' only in the first component
of the grading, and improves on this successively.

The grading splits the BRST charge into three parts $Q=Q_0+Q_1+Q_2$, with
$[Q_0]=(1,0)$, $[Q_1]=(\frac 1 2,\frac 1 2)$ and $[Q_2]=(0,1)$.
The operators $Q_0$, $Q_1$ and $Q_2$, map $\ca_{(m,n)}$ to $\ca_{(m+1,n)}$,
$\ca_{(m+\frac 1 2 ,n+ \frac 1 2)}$ and $\ca_{(m,n+1)}$ respectively:
\bea
Q_0&=&-\frac{\k}{8\p i x}\oint str c e_-\nonu
Q_1&=&-\frac{\k}{8\p i x}\oint str c\left[\t,e_-\right].\label{qsqr}
\eea
Nilpotency of $Q$ implies that
$Q_0^2=Q_2^2=\{Q_0,Q_1\}=\{Q_1,Q_2\}=Q_1^2+\{Q_0,Q_2\}=0$, but \be
Q_1^2=-\{Q_0,Q_2\}=\frac{\k}{32 \p i x}\oint str\left\{ c \left[
\P_{1/2}c,e_-\right] \right\}\label{Q12}
\ee
does not vanish - this is the obstruction to the existence of a double complex.
The action of $Q_0$, $Q_1$ and $Q_2$ on the basic fields is given by
{\small
\be
\begin{array}{rcclrcclrccl}
Q_0\,: &  b & \rightarrow & -\frac \k 2 e_- &
Q_1\,: &  b & \rightarrow & -\frac \k 2 [\t,e_-] &
Q_2\,: &  b & \rightarrow & \P_-\hat{J}_z\\
 &c &\rightarrow&  0&
 &c &\rightarrow & 0&
 &c &\rightarrow & \frac 1 2 cc  \\
 &\hat{J}_z& \rightarrow&-\frac \k 4 [e_-,c]&
 &\hat{J}_z& \rightarrow&-\frac \k 4 [[\t,e_-],c]&
 &\hat{J}_z& \rightarrow&\frac 1 2 \vec{[}c,\overline{\P}_-\hat{J}_z]+\frac \k
4 \del c \\
 & & & & & & & & & & &-\frac 1 2 [\P_-(t^A),[\P_+(t_A),\del c ]] \\
 &\t &\rightarrow&  0&
 &\t &\rightarrow & \frac 1 2 \P_{+1/2} c&
 &\t &\rightarrow & 0,\end{array}\label{qonf}
\ee }
\noindent
where
$\overline{\P}_-\equiv {\bf 1}-\P_-$ and $\vec{[}A,B]$ stands for
\be
\vec{[}X,Y]=(-)^{(AB)}\left(X^AY^B \right)f_{AB}{}^Ct_C,
\ee
where $\left(X^AY^B \right)$ is a regularized product.
The BRST charge $Q$ acts as a derivation on a regularized product of fields.

To exploit the double grading, we need one more preparation. For a fixed
ghost number $k$, the sequence of gradings $(m,k-m)$ is unbounded because the
first component is negative for some currents and for the $b$-field.
The application of spectral sequence techniques requires this set
to be finite. This is remedied by
noticing that the subcomplex $\ca^{(1)}$, generated by $\{ b,
\P_-\hat{J}_z-\frac \k 2 [ \t,e_-]\}$ has a trivial cohomology
$H^*(\ca^{(1)};Q)={\bf C}$. From this and eq. (\ref{qonf}) we find that we can
as well compute the cohomology of the reduced complex $\wha=\ca / \ca^{(1)}$,
generated by $\{\overline{\P}_-\hat{J}_z,\t,c\}$ as
$H^*(\ca)=H^*(\wha)$.  The OPEs close on the reduced complex. The double
grading on $\ca$ induces a double grading on $\wha$.
The filtration
$\wha^m$, $m\in\frac 1 2 {\bf Z}$ of $\wha$:
\be
\wha^m\equiv \bigoplus_{k\in\frac 1 2 {\bf Z}}\bigoplus_{l\geq m}\wha_{(k,l)}.
\ee
now leads to  a spectral sequence $(E_r,d_r)$, $r\geq1$, converging
to $H^*(\wha;Q)$.  Each term in the sequence is the cohomology of the previous
term with a derivation that represents the effective action of the
BRST operator at that level: $E_r=H^*(E_{r-1};d_{r-1})$. To start with,
this means that one neglects terms that are one half unit lower in the
first grading component. The first term in the
sequence is then $E_0=\wha$, $d_0=Q_0$.

The next term is $E_1= H^*(\wha ;Q_0)$. The derivation operator at this level
will act like $Q$ up to terms that have the first gradation component
at least one unit lower.
Thus on $E_1$, the $Q_0$ cohomology classes, it acts like $Q_1$, i.e.
if $\F\in\wha, [\F]\in E_1, d_1[\F]=[Q_1]\F$.
Note that although $Q_1^2 \ne 0$, nevertheless $d_1^2=0$,
since $Q_1^2=-Q_0Q_2-Q_2Q_0$ implies that on
the $Q_0$ cohomology classes it always results in the trivial class.

The next term is then $E_2=H^*(E_1 ;d_1)$. The $d_2$ derivation can be
computed as follows. Let $\F$ represent some class $[[\F]] \in E_2$.
 Then $Q_0(\F)=0$, and $Q_1(\F)=Q_0(\J)$
 for some $\J\in\wha$. The action of
$d_2$ is $d_2[[\F]]=[[\F_2 \equiv Q_2(\F)-Q_1(\J)]]$. One may
check that this is a proper operation in $E_2$, since
$Q_0\F_2=0$ and $Q_1\F_2=Q_0(Q_2\J)$. It is equally trivial to verify that
$d_2^2=0$. The $d_2$-cohomology  gives the next term in the sequence,
and so on, until the sequence stabilizes (i.e. the derivation operator
 vanishes).

Applying this to $H^*(\wha ;Q)$, using eq. (\ref{qonf}), one finds that
$E_1$ is represented by\footnote{From now on $\hat{J}_z$ stands for
$\overline{\P}_-\hat{J}_z$}
\be
E_1\simeq\wha \left[\P_{\ker ad e_+}\hat{J}_z \right]\otimes\wha \left[
\t\right]\otimes\wha \left[ \P_{\frac 1 2 }c\right],
\ee
where we denoted the subalgebra of $\wha$ generated by $\F$ by $\wha \left[\F
\right]$. Using eq. (\ref{qonf}) again, one finds that $E_2$ is represented by
\be
E_2\simeq\wha \left[\P_{\ker ad e_+}\left(\hat{J}_z +\frac\k
4[\t,[e_-,\t]]\right)\right],
\ee
and one has explicitely
\be
Q_1\P_{\ker ad e_+}\left(\hat{J}_z +\frac\k 4[\t,[e_-,\t]]\right)=Q_0\P_{\ker
ad
e_+}\left[ \t,\hat{J}_z\right].\label{explf}
\ee

It turns out that $E_2$ has only ghost number zero elements.
Therefore, $d_2$ is actually the zero map and the sequence has already
stabilized at the previous level.
This gives the main result:
\be
H^*(\ca;Q)\simeq E_2=H^*(H^*(\wha,Q_0),d_1).
\ee

Having established the cohomology of $Q$, we now turn to the explicit
construction of its generators. The cohomology is generated by
$\hat{T}\equiv\sum_{j,\a_j}\hat{T}^{(j,\a_j)}t_{(jj;\a_j)} \in \ker ad \, e_+$
and $\hat{T}^{(j,\a_j)}$ has the form
\be
\hat{T}^{(j,\a_j)}=\sum_{r=0}^{2j}\hat{T}^{(j,\a_j)}_r,
\ee
where $\hat{T}^{(j,\a_j)}_r$ has grading $(j-\frac r 2 , -j+\frac r 2)$. From
the previous discussion, we know that the leading term $\hat{T}^{(j,\a_j)}_0$
is of the form
\be
\hat{T}^{(j,\a_j)}_0=C^j\left\{ \hat{J}_z^{(jj;\a_j)}+\frac \k 4
\sum_{\a_0}\d_{j,0}\d_{\a_j,\a_0} [\t ,[e_-,\t]]^{(00;\a_0)}
\right\}\label{lterm}
\ee
where the normalization constant $C$ will be fixed later on. The remaining
terms are recursively determined by a generalized tic-tac-toe construction. We
have
\be
Q_0\hat{T}^{(j,\a_j)}_r=-Q_1 \hat{T}^{(j,\a_j)}_{r-1} - Q_2
\hat{T}^{(j,\a_j)}_{r-2},
\ee
where $\hat{T}^{(j,\a_j)}_r=0$ for $r<0$ or $r>2j$. This determines
$\hat{T}^{(j,\a_j)}_r$ modulo, the addition of an arbitrary functional of $\t$.
As a functional of $\t$ only has grading $(0,0)$, it can only appear in
$\hat{T}^{(j,\a_j)}_{2j}$. It gets determined by the final recursion relation:
$Q_1 \hat{T}^{(j,\a_j)}_{2j} + Q_2 \hat{T}^{(j,\a_j)}_{2j-1}=0$.
As an example we compute $\hat{T}^{(0,\a_0)}$ and $\hat{T}^{(1/2,\a_{1/2})}$.
{}From eqs. (\ref{qonf}) and (\ref{explf}), one immediately gets
\be
\hat{T}^{(0,\a_0)}=\hat{T}^{(0,\a_0)}_0=
\hat{J}_z^{(00;\a_0)}+\frac \k 4 [\t ,[e_-,\t]]^{(00;\a_0)}.
\ee
For $j=1/2$, one has $\hat{T}^{(1/2,\a_{1/2})}_0= \sqrt{C}
\hat{J}_z^{(1/2,1/2;\a_{1/2})}$. From eq. (\ref{explf}), it follows that
$\hat{T}^{(1/2,\a_{1/2})}_1=-\sqrt{C}[\t,\hat{J}_z]^{(1/2,1/2;\a_{1/2})} +f
(\t)$. The unknown function $f$ gets determined by the next recursion relation
$Q_2\hat{T}^{(1/2,\a_{1/2})}_0 + Q_1 \hat{T}^{(1/2,\a_{1/2})}_1 = 0$: $f (\t) =
\sqrt{C}(\frac \k 6 [\t,[\t,[\t , e_-]]]-\frac \k 2 \del \t -
[\P_-(t^A),[\P_+(t_A),\del \t ]])$. We have that $\hat{T}^{(1/2,\a_{1/2})}=
\hat{T}^{(1/2,\a_{1/2})}_0 +\hat{T}^{(1/2,\a_{1/2})}_1$.

The explicit form of all generators $\hat{T}^{(j,\a_j)}$ must obviously be
computed on a case by case basis. A useful tool for this is the fact that one
can always construct $\hat{J}'_z\in\P_{\geq 1}\bar{g}$ such that
$[e_-,\hat{J}'_z]=\hat{J}_z-\P_{\ker ad e_+}\hat{J}_z$. This implies that $c$
can
be written as $c= -\frac 4 \k Q_0(\hat{J}'_z)+2Q_1(\t)$. Note also that
$Q_2(\hat{J}_z)$ can be rewritten as $Q_2(\hat{J}_z)=\frac 1 4
\vec{[}c,\overline{\P}_-\hat{J}_z]  - \frac 1 4
\vec{[}\overline{\P}_-\hat{J}_z,c]+\frac {\k+\tilde{h}}
{4} \del c +\frac 1 4 [\P_0(t^A),[\P_0(t_A),\del c ]]$.

The energy-momentum tensor $\hat{T}^{\rm EM}\equiv \hat{T}^{(1,0)}$ itself can
be computed. It is given by
\bea
\hat{T}^{\rm EM}&=&\frac{\k}{x(\k+\tilde{h})}\bigg( str \left\{\hat{J}_z e_-
\right\} + str \left\{ [\t,e_- ] \hat{J}_z\right\} +\frac 1 \k str
\left\{\P_0(\hat{J}_z)\P_0(\hat{J}_z) \right\} +\frac
{\k+\tilde{h}}{\k}str\left\{ e_-\del\hat{J}'_z\right\} \nonu
&& + \frac {1}{\k}str\left\{ \left[\P_0 (t^A),\left[\P_0 (t_A) ,\del\hat{J}'_z
\right]\right]e_-\right\}-\frac{\k+\tilde{h}}{4}str \left\{ [\t,e_-]\del\t
\right\}\bigg).
\eea
The first term is $\hat{T}^{(1,0)}_0$, the second term $\hat{T}^{(1,0)}_1$ and
the remainder forms $\hat{T}^{(1,0)}_2$.

It is not hard to verify that $\hat{T}^{\rm IMP}$,
\bea
\hat{T}^{\rm IMP}
&\equiv&\frac{1}{x(\k+\tilde{h})} str J_zJ_z -\frac{1}{8xy}str e_0 \del J_z
-\frac{\k}{4 x}tr\left( [\t,e_-]\del\t\right)\nonu
&&+\frac{1}{4x}strb[e_0,\del c]-\frac {1}{2x}str b\del c+\frac{1}{4x}str
\del b
[e_0,c]
,\label{Timp}
\eea
is also BRST invariant. It differs from $\hat{T}^{\rm EM}$ by a BRST exact term
$Q\left(-\frac{2}{x(\k+\tilde{h})}str b\hat{J}_z+\cdots \right)$. Eq.
(\ref{Timp}) is particularly useful to compute the central extension
\cite{kris2}. One finds:
\be
c=\frac 1 2 c_{\rm crit} - \frac{(d_B-d_F)\tilde{h}}{\k+\tilde{h}} - 6 y (\k
+\tilde{h}),\label{cpretty}
\ee
where $d_B$, $d_F$ respectively, is the number of bosonic, fermionic
respectively, generators of $\bar{g}$, $c_{\rm crit}$ is the critical value of
the central charge for the
extension of the Virasoro algebra under consideration:
\be
c_{\rm crit}=\sum_{j,\a_j}(-)^{(\a_j)}(12 j^2+12j+2).\label{ccrit}
\ee
and $y$ is the index of embedding, which in the case $\tilde{h}\neq 0$ is given
by
\be
y=\frac{1}{3\tilde{h}}\sum_{j\a_j}(-)^{(\a_j)}j(j+1)(2j+1)\label{indexem}
\ee
and $(-)^{(\a_j)}=+1$ ($-1$) if $t_{(jm,\a_j)}$ is bosonic (fermionic).
Note that the requirement that $\hat{T}^{\rm EM}$ generates the Virasoro
algebra in the standard normalization fixes the normalization constant $C$ to
be
\be
C=\frac{4y\k}{\sqrt{2}(\k+\tilde{h})}.
\ee
Knowing the leading term of the currents, eq. (\ref{lterm}), we find that the
conformal dimension of $\hat{T}^{(j,\a_j)}$ is given by $j+1$.

We already observed that the OPEs close on $\wha$ and preserve the grading.
{}From this we deduce that the OPEs of the generators $\hat{T}$ close modulo
BRST exact terms. However, we work on the reduced complex $\wha$, which has no
states of negative ghostnumber. So there are no BRST exact terms at ghost
number zero. We conclude that the OPEs of the generators $\hat{T}$ close.

The quantum Miura transformation for the generators of the extended Virasoro
algebra is also easily obtained. As the OPEs preserve the grading, the  part of
$\hat{T}$ with grading $(0,0)$, $\hat{T}_{(0,0)}$, closes among themselves. So
the map $\hat{T}\rightarrow\hat{T}_{(0,0)}$ is an algebra homomorphism. In
order to prove that the map $\hat{T}\rightarrow\hat{T}_{(0,0)}$ is an algebra
isomorphism, we have to show that each generator of the extended Virasoro
algebra has a non-vanishing component of grading $(0,0)$. This is shown
following a reasoning similar to the one in \cite{tjin1}. Consider the mirror
of the spectral sequence, {\it i.e.} the one associated to the filtration
\be
\wha'^m\equiv \bigoplus_{l\in\frac 1 2 {\bf Z}}\bigoplus_{k\geq m}\wha_{(k,l)}.
\ee
We already know that $E'_\infty$ for this spectral sequence vanishes unless the
ghostnumber is zero. Using eq. (\ref{qonf}), one shows that $E_1=H^*(\wha ;
Q_2)$ is only non-vanishing at grading $(\frac m 2 ,\frac m 2)$, $m\geq 0$.
This implies that $E_\infty$ is only non-vanishing at grading $(0,0)$. This
proves that
$\hat{T}\rightarrow\hat{T}_{(0,0)}$ is indeed an algebra isomorphism.

Finally, the method of working we described is particularly useful to compute
the effective action in the light-cone gauge, $W[\check{T}]$, of the
corresponding gravity theory, \cite{kris2}. The effective action is defined by
\be
\exp -W[\check{T}]=\int [\d g g^{-1}][d\t][d A_{\bz}][d\m ]\left(
\mbox{Vol}\left( \P_+\bar{g} \right) \right)^{-1}\exp-\left( {\cal S}_2
-\frac{1}{4\p x y} \int str \m\check{T}\right).\label{okok2}
\ee
The effective action is most easily computed by making a different gauge
choice. We make a canonical transformation in eq. (\ref{bvbv}) which
interchanges fields and anti-fields for $\{\t,\t^*\}$ and $\{ \P_+[e_+,J_z],
\P_-[e_-,J^*_z]\}$. This corresponds to choosing the gauge
$\t=\P_+[e_+,J_z]=0$.  We find
\be
W[\check{T}]=\k_c S_-[g]\label{endresult}
\ee
where $\k_c=\k+2\tilde{h}$ and we used $[\d g g^{-1}]=[dJ_z]\exp \left(
-2\tilde{h} S^-[g]\right)$.
{}From eq. (\ref{cpretty}) we get the level as a function of the central
charge:
\be
12 y \k_c=12 y\tilde{h}-\left(c-\frac 1 2 c_{\rm crit}\right)-
\sqrt{\left(c-\frac 1 2 c_{\rm crit}\right)^2- 24 (d_B-d_F)
\tilde{h}y}\label{vv2}
\ee
Eq. (\ref{vv2}) provides an all-order expression for the
coupling constant renormalization.
The WZW model in eq. (\ref{endresult}) is constrained by
\be
\del g g^{-1}+\frac{1}{4xy}str \left\{\P_{\rm NA}\left(\del g
g^{-1}\right))\P_{\rm NA}\left(\del g g^{-1}\right)\right\}e_+=e_-+\frac{1}{\k
+\tilde{h}} \sum_{j,\a_j}
\frac{1}{2^{\frac 3 2 j-1}y^{j}}
\check{T}^{(j\a_j)} t_{(jj,\a_j)},
\ee
where  $\P_{\rm NA}\bar{g}$ is the projection on the centralizer of
$\slt$ in $\bar{g}$. We also used $J_z=\frac{\a_\k}{2}\del g g^{-1}$ with
$\a_\k=\k+\tilde{h}$. For a detailed discussion on the value of $\a_\k$, see
\cite{kris1} and in particular \cite{ruud}.

The strategy developed in this paper has numerous applications. In particular,
the representation theory of the extended Virasoro algebra should be closely
related to the representation theory of the corresponding affine Lie algebra.

The approach followed here was closely related to 2D gravity in the light-cone
gauge. In order to study {\it e.g.} non-critical strings based on some extended
Virasoro algebra, one needs a covariant formulation. It is  clear from eq.
(\ref{vv2}) that a minimal $(p,q)$ matter sector is realized in terms of a
gauged WZW model with level $\k_M=p/q -\tilde{h}$ and in order to cancel the
conformal anomaly we need a gauge sector based on a WZW model with level
$\k_M=-p/q -\tilde{h}$. The gauge invariant coupling between the matter and the
gauge sector is performed in a way somewhat similar to the path followed in
\cite{jacob}. The main problem is to construct the analogue of eq.
(\ref{invact})
in which both the left and the right moving extended Virasoro are coupled to
sources in a gauge invariant way, {\it i.e.} we need an action of the form
${\cal S}_2={\cal S}_1+\frac{1}{4\p x y} \int str ( \m T+\bar{\m}
\overline{T})$. This can be achieved by following a path inspired by the
methods developed in \cite{first}. Details about this will be reported on
elsewhere \cite{alex2}.


\end{document}